\documentclass[12pt]{article}

\usepackage{cite}
\usepackage{epsfig}
\usepackage{amsmath}

\def\mathswitch#1{\relax\ifmmode#1\else$#1$\fi}
\def\mathswitchr#1{\relax\ifmmode{\mathrm{#1}}\else$\mathrm{#1}$\fi}
\newcommand{\PW}{\mathswitchr W}

\newcommand{\Pt}{\mathswitchr t}

\newcommand{\mt}{\mathswitch {m_\Pt}}

\newcommand{\scrs}{{}}
\newcommand{\sw}{\mathswitch {s_{\scrs\PW}}}
\newcommand{\cw}{\mathswitch {c_{\scrs\PW}}}

\newcommand{\gev}{\,\, \mathrm{GeV}}

\newcommand{\SLASH}[2]{\makebox[#2ex][l]{$#1$}/}

\newcommand{\Eslash}{\SLASH{E}{.5}\,}

\newcommand{\anc}{\rule{0mm}{0mm}}
\newcommand{\lesim}{\,\raisebox{-.1ex}{$_{\textstyle <}\atop^{\textstyle\sim}$}\,}

\newcommand{\mycaption}[1]{\caption{\sl #1}}

\begin{document}
\thispagestyle{empty}

\def\thefootnote{\fnsymbol{footnote}}

\begin{flushright}
ZH-TH 20/06
\end{flushright}

\vspace{1cm}

\begin{center}

{\Large\sc {\bf Phenomenology of Mirror Fermions in the Littlest Higgs Model 
with T-Parity}}
\\[3.5em]
{\large\sc 
A.~Freitas
and
D.~Wyler
}

\vspace*{1cm}

{\sl
Institut f\"ur Theoretische Physik,
        Universit\"at Z\"urich, \\ Winterthurerstrasse 190, CH-8057
        Z\"urich, Switzerland
}

\end{center}

\vspace*{2.5cm}

\begin{abstract}

Little Higgs models are an interesting alternative to explain electroweak
symmetry breaking without fine-tuning.  Supplemented with a discrete symmetry
(T-parity)  constraints  from electroweak precision data are naturally evaded
and also a viable dark matter candidate is obtained. T-parity implies the
existence of  new (mirror) fermions in addition to the heavy gauge bosons of
the little Higgs models.  In this paper we consider the effects of the mirror
fermions on the phenomenology of the littlest Higgs model with T-parity at the
LHC. We study  the most promising production channels and decay chains for the
new particles. We find that the mirror fermions have a large impact on the
magnitude of signal rates and on the new physics signatures. Realistic
background estimates are given.  

\end{abstract}

\def\thefootnote{\arabic{footnote}}
\setcounter{page}{0}
\setcounter{footnote}{0}

\newpage


\section{Introduction}

A simple doublet scalar field yields a perfectly appropriate gauge symmetry
breaking pattern in the  Standard Model (SM). On the other hand, its
theoretical shortcomings, such as quadratic  divergencies (hierarchy problem)
or the triviality of a $\phi^4$ theory suggest that it is  embedded in a
larger scheme. On the other hand,  electroweak precision data suggests that up
to a scale of about 10 TeV, no new strong interaction is present,  indicating
that indeed some form of a scalar interaction is required, including the 
possibility that the (Higgs) scalars  are composite. A much investigated
option to solve the hierarchy problem are supersymmetric theories; but also
models with extra dimensions have been considered. 

Recently, an alternative known as the \emph{little Higgs mechanism}
\cite{little}, has been proposed  where the smallness of the electroweak scale
is assured by interpreting the Higgs as a (Pseudo) Goldstone particle of a
symmetry breakdown at a scale $f$. The new gauge bosons
and  partners of the top quark  with a mass of order $f$  cancel the one-loop
quadratic corrections to the Higgs mass from Standard Model (SM) particles. 
A very appealing implementation of the little Higgs concept is the
\emph{littlest} Higgs model \cite{littlest}, which extends the SM by a
minimal number of gauge bosons and fermions.

However, even though these new particles 
are weakly coupled, electroweak precision data requires
$f$ to be above 5 TeV \cite{lhew}. On the other hand a scale as low as 1 TeV
is required to avoid fine-tuning of the Higgs mass.

A discrete symmetry, called \emph{T-parity} \cite{LHT, wudka} circumvents these 
problems. It forbids all
tree-level contributions of the new heavy degrees of freedom to electroweak
precision observables. The SM fields are T-even, while the new TeV-scale
particles are odd. Therefore, the new particles can only be generated in
pairs which is reminiscent of R-parity in supersymmetric theories.
Besides satisfying the electroweak constraints, T-parity also has the
interesting consequence that the lightest T-odd particle is stable and, if
neutral, a good candidate for cold dark matter.

One of the prime objectives of the next generation of colliders, especially the
Large Hadron Collider (LHC), is to unravel the physical mechanism of
electroweak symmetry breaking.  A first study of the phenomenology of the
littlest Higgs model with T-parity (LHT) at the LHC presented in
Ref.~\cite{hubmed} yielded attractive production rates for the new
particles.  The collider phenomenology of little Higgs theories with T-parity
has some similarities to supersymmetry. In particular, it involves signatures
with missing energy originating from the neutral lightest T-odd particle (LTP),
which escapes detection. On the other hand, little Higgs theories usually do
not involve a T-odd partner of the gluon, which leads to smaller new physics
cross-sections at the LHC compared to supersymmetry.

In this paper, the phenomenology of the littlest Higgs model with T-parity is
revisited and in particular the important role of  the T-odd fermions
(\emph{mirror fermions}) is stressed. Backgrounds from SM sources are
investigated in order to arrive at realistic estimates for the observability
of the new physics signals of the LHT model. After reviewing the LHT model in
section \ref{sc:model}, the production processes and signatures for
various T-odd particles are studied in detail in sections \ref{sc:lhc} and
\ref{sc:sign}. Finally, the conclusions are given in section \ref{sc:concl}.


\section{The model}
\label{sc:model}

The littlest Higgs model is based on a coset SU(5)/SO(5), i.~e. a global SU(5)
symmetry that is explicitly broken down to a SO(5) group, with a $[SU(2) \times
U(1)]^2$ subgroup of SO(5) being gauged \cite{littlest}. The Goldstone modes of
the broken SU(5) are implemented in a non-linear sigma model with a breaking
scale $f$. Some of the Goldstone modes only become massive when {\it both} gauge
subgroups are broken. As a consequence of this simultaneous symmetry breaking
one-loop quadratic divergencies to the Higgs mass are naturally avoided and their  mass
get corrections at most at the two-loop level.

The littlest Higgs model can be supplemented by a discrete $Z_2$ called T-parity
\cite{LHT}, with SM
particles being even ($T=+1$), and non-SM particles odd ($T=-1$) under this
symmetry. The terms of the non-linear sigma generate masses
of order $f$ for the T-odd particles. Rather than giving the whole construction of the model 
we consider, we refer to \cite{hubmed} for a somewhat.
detailed description. 

The gauge bosons are formed from the gauge bosons of the two SU(2) and U(1)
groups, which in the following are indicated by subscripts 1 and 2,
respectively:
\begin{align}
W_L^a &= \tfrac{1}{\sqrt{2}}(W_1^a + W_2^a), && \text{(T-even)} \\
B_L &= \tfrac{1}{\sqrt{2}}(B_1 + B_2), 
\intertext{with masses from usual electroweak symmetry breaking, and}
W_H^a &= \tfrac{1}{\sqrt{2}}(W_1^a - W_2^a), && \text{(T-odd)} \\
B_H &= \tfrac{1}{\sqrt{2}}(B_1 - B_2), 
\end{align}
with masses of order $f$ generated from the kinetic term of the non-linear
sigma model. After electroweak symmetry breaking, the light gauge bosons mix to
form the usual physical states of the SM, $A_L = \cw B_L - \sw W_L^3$, $Z_L =
\sw B_L + \cw W_L^3$ and $W_L^\pm = (W_L^1 \mp W_L^2)/\sqrt{2}$. Similarly, a
small mixing of order ${\cal O}(v^2/f^2)$ is introduced between $B_H$ and $W_H^0 \equiv
Z_H$ through electroweak symmetry breaking. In this work, this mixing and all
other terms of order ${\cal O}(v^2/f^2)$ will be consistently neglected. In this
case, the masses of the T-odd gauge bosons are
\begin{equation}
M_{W_H^\pm} = M_{Z_H} = gf, \qquad 
M_{B_H} = \frac{g'}{\sqrt{5}}f,
\end{equation}
with $W_H^\pm = (W_H^1 \mp W_H^2)/\sqrt{2}$.
The $B_H$ is always lighter than the other T-odd gauge bosons and thus a good
candidate for the LTP (lightest T-odd particle) and dark matter.

T-parity also requires a doubling of the fermion sector associating to each 
T-even (SM) fermion a T-odd fermion. $F_H$ (mirror fermion).
 These 'partners' get  masses
\cite{mirrormass}
\begin{equation}
m_{f_{H,i}} = \sqrt{2} \kappa_{i} f,
\end{equation}
where the Yukawa couplings $\kappa$ can in general depend on the fermion species
$i$. The $\kappa$ can also generate flavor changing
interactions, but this will not be studied in this paper.

The implementation of the mass terms for the mirror fermions also introduces T-odd
SU(2)-singlet fermions, which
may receive large masses and do not mix with the SU(2)-doublets $f_H$.
Here it is therefore assumed that these extra singlet fermions have large masses
and decouple from phenomenology at the LHC.

The top sector requires an additional T-even fermion $t'_+$ and one T-odd
fermion $t'_-$ to cancel quadratic divergencies to the Higgs mass. Their masses
are
\begin{equation}
\mt = \frac{\lambda_1\lambda_2v}{\sqrt{\lambda_1^2 + \lambda_2^2}}, \qquad
m_{t'_+} = \sqrt{\lambda_1^2 + \lambda_2^2} \, f, \qquad
m_{t'_-} = \lambda_2 f.
\end{equation}
The Yukawa couplings $\lambda_{1,2}$ are constrained by the top mass $\mt$, but
one has the freedom to choose
\begin{equation}
s_\lambda \equiv \frac{\lambda_2}{\sqrt{\lambda_1^2 + \lambda_2^2}} =
\frac{m_{t'_-}}{m_{t'_+}}.
\end{equation}
The decay modes of the T-odd gauge bosons and mirror fermions are summarized in
Tab.~\ref{tab:decays} for $f = 1$ TeV, degenerate $\kappa_f = 0.5$ and
$s_\lambda = 1/\sqrt{2}$.
\renewcommand{\arraystretch}{1.2}
\begin{table}[tp]
\begin{center}
\begin{tabular*}{\textwidth}{c@{\extracolsep\fill}ccr@{\extracolsep{0pt}\:}l@{\hspace{2em}}l}
\hline
Particle & Mass $m$ [GeV] & Width $\Gamma$ [GeV]
          & \multicolumn{3}{c}{Decay modes} \\
\hline 
$W_H^\pm$ & $647$  & $0.056$ &
        $W_H^\pm$ & $\to W^\pm \, B_H$ & 100\% \\
$Z_H$ & $647$  & $0.051$ &
        $Z_H$ & $\to h^0 \, B_H$ & 100\% \\
$B_H$ & $154$  & --- & \multicolumn{2}{c}{---} & \\
\hline
$u_H$ & $705$  & $0.36$ &
        $u_H$ & $\to B_H \, u$ & 50\% \\
        &&&& $\to W_H^+ \, d$ & 33\% \\
        &&&& $\to Z_H \, u$ & 17\% \\
$d_H$ & $705$  & $0.36$ &
        $d_H$ & $\to B_H \, d$ & 50\% \\
        &&&& $\to W_H^+ \, u$ & 33\% \\
        &&&& $\to Z_H \, d$ & 17\% \\
$t'_-$ & $1000$ & 1.25 &
	$t'_-$ & $\to B_H \, t$ & 100\% \\
\hline
\end{tabular*}
\end{center}
\vspace{-1em}
\mycaption{Tree-level masses, widths and main branching ratios of the
T-odd states at Born level for $f = 1$ TeV, $\kappa_f = 0.5$ and
$s_\lambda = 1/\sqrt{2}$. Corrections of order ${\cal O}(v^2/f^2)$ are
neglected.}
\label{tab:decays}
\end{table}


\section{T-odd  production at LHC}
\label{sc:lhc}

In this section we consider the (pair) production of the various T-odd particles at the
LHC. Our work completes the previous studies  in Ref.~\cite{hubmed} and stresses
the role of the T-odd fermions.


\subsection{Heavy gauge bosons}

\begin{figure}[tb]
\psfig{figure=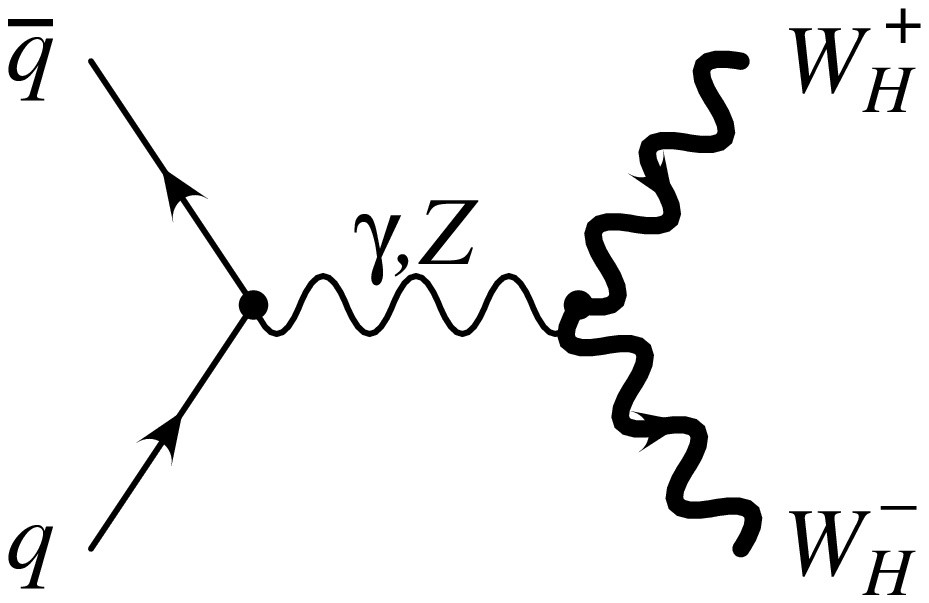, width=5cm}
\psfig{figure=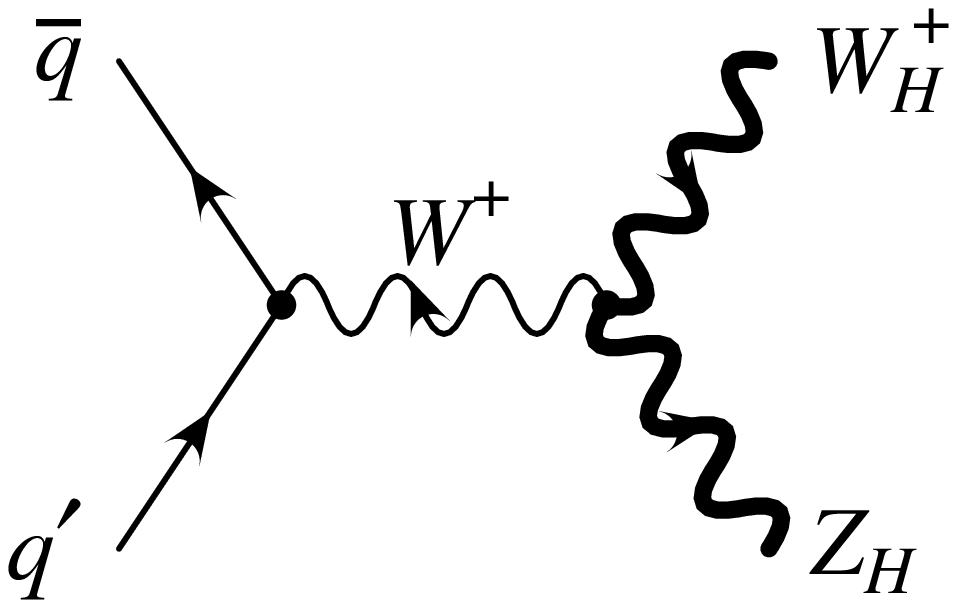, width=5cm}
\psfig{figure=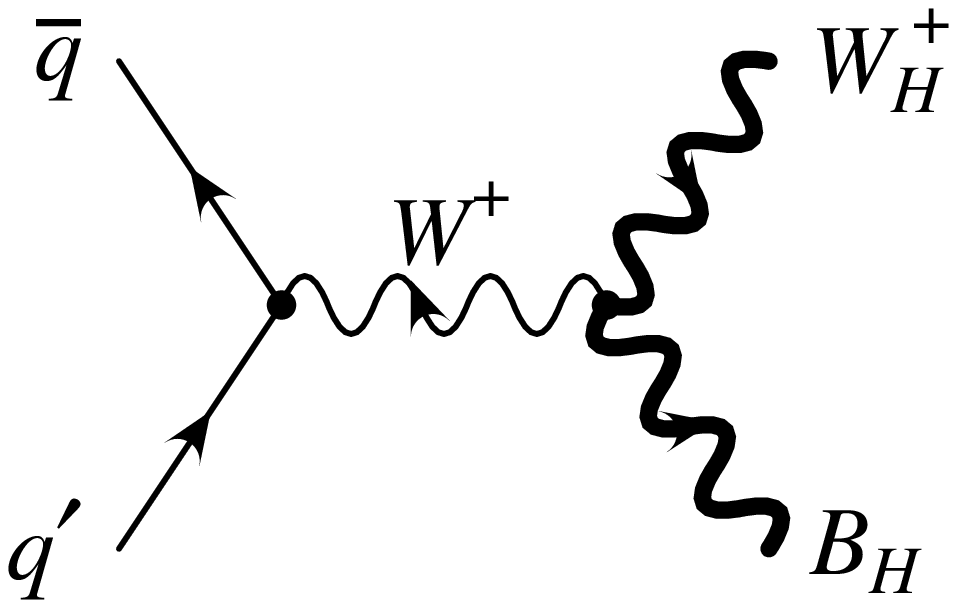, width=5cm}
\\
\anc\hspace{1.8cm}(a)\hspace{4.7cm}(b)\hspace{4.7cm}(c)\\[1.5em]
\psfig{figure=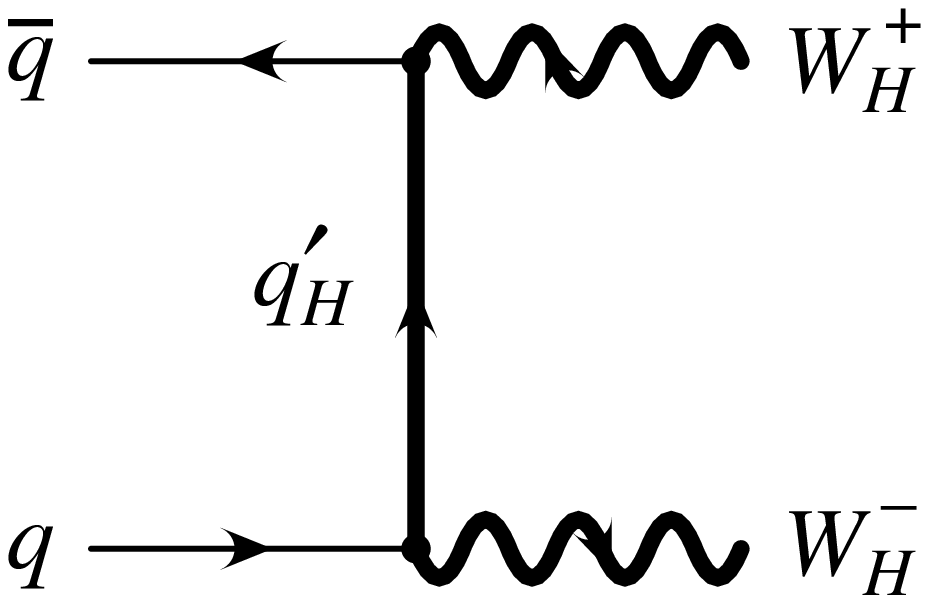, width=5cm}
\psfig{figure=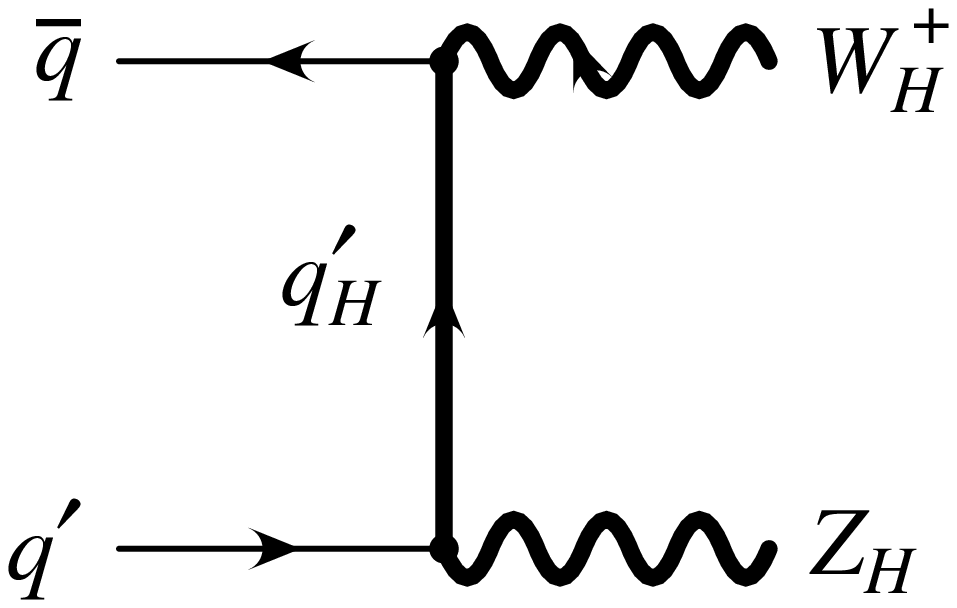, width=5cm}
\psfig{figure=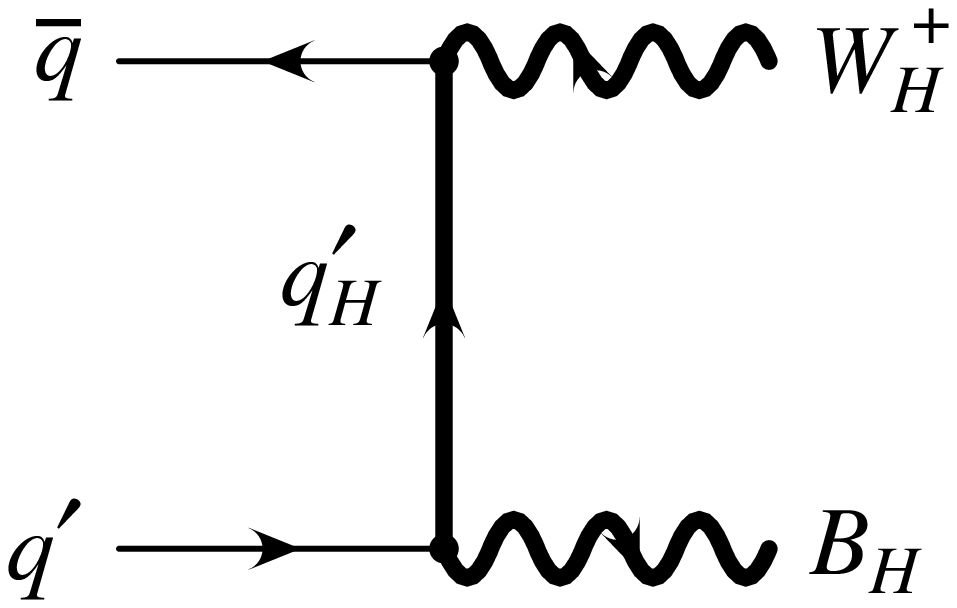, width=5cm}
\\
\anc\hspace{1.8cm}(d)\hspace{4.7cm}(e)\hspace{4.7cm}(f)
\mycaption{Production diagrams for T-odd gauge bosons at LHC. 
Thick lines indicate T-odd propagators.}
\label{fig:diav}
\end{figure}

In Ref.~\cite{hubmed}, the production of heavy T-odd gauge bosons was computed
including the s-channel gauge boson contributions only,
Fig.~\ref{fig:diav}~(a--c). It was found that the cross-sections can be sizeable,
of the order ${\cal O}$(0.1 pb), for $f < 1$ TeV.
However, for perturbative
Yukawa couplings $\kappa_f \lesim 1$, the mirror fermions cannot be much heavier
than the heavy gauge bosons, and thus never decouple from the production
process. Thus the t-channel mirror fermion exchange,
Fig.~\ref{fig:diav}~(d--f), is also important. 

To study this effect of the mirror fermions for heavy gauge boson production,
the Feynman rules for the littlest Higgs model with T-parity
\cite{hanetal,hubmed,hubpaz} have been implemented in CompHEP \cite{CompHEP}.
The package CompHEP was then used to generate numerical results for production
cross-sections and decay modes of the T-odd heavy gauge bosons.

\begin{figure}[tb]
\anc\hspace{1.9cm}$pp \to W_H^+ W_H^-$
\anc\hspace{1.5cm}$pp \to W_H^+ Z_H^{}$
\anc\hspace{1.65cm}$pp \to W_H^+ B_H^{}$\\[-3em]
\anc\hspace{-4ex}%
\psfig{figure=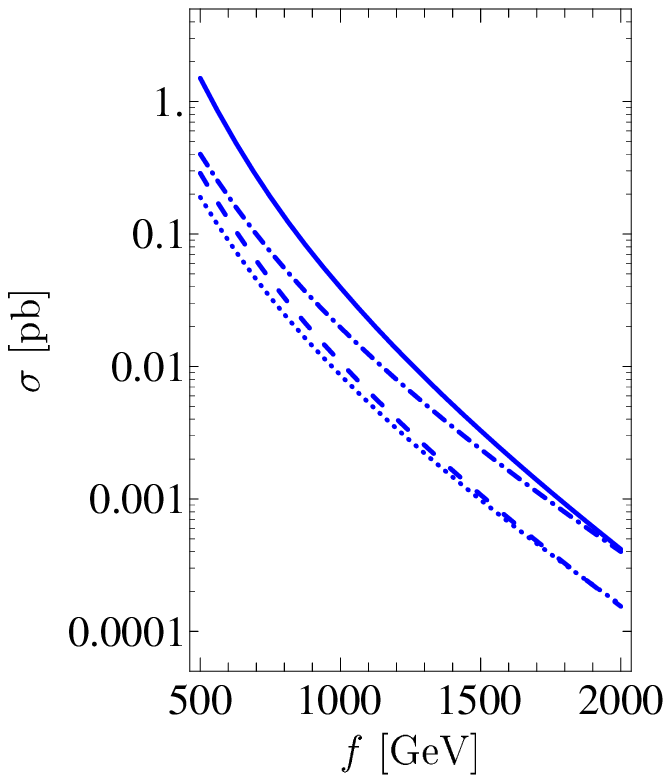, height=9.5cm}\hspace{-2ex}%
\psfig{figure=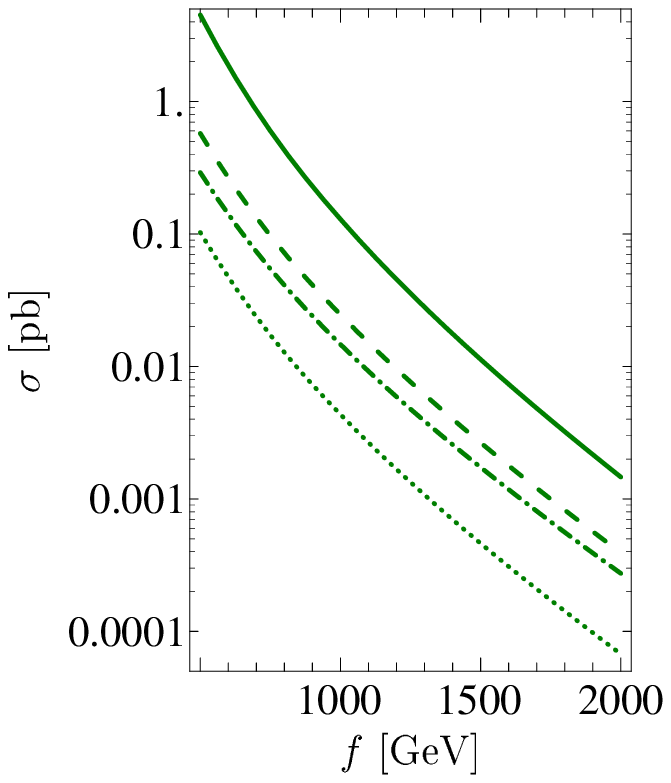, height=9.5cm, 
	bb=80 382 221 682, clip=true}\hspace{-2ex}%
\psfig{figure=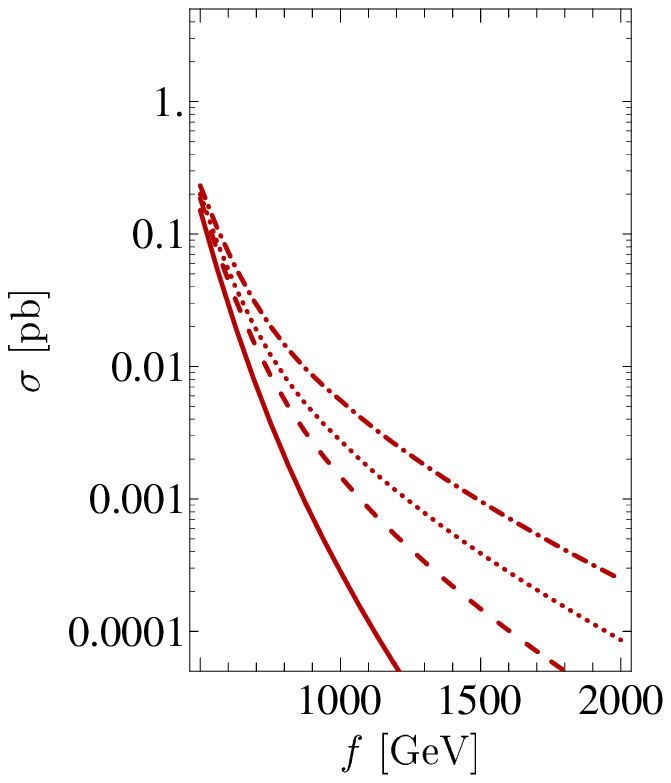, height=9.5cm,
	bb=80 382 221 682, clip=true}
\begin{minipage}[b]{1.5cm}
\rule{10mm}{.3mm} $\kappa = \infty$\\[1ex]
\rule{2mm}{.3mm}\rule{2mm}{0mm}\rule{2mm}{.3mm}\rule{2mm}{0mm}\rule{2mm}{.3mm}
$\kappa = 1$\\[1ex]
\rule{.5mm}{.3mm}\rule{.5mm}{0mm}%
\rule{.5mm}{.3mm}\rule{.5mm}{0mm}%
\rule{.5mm}{.3mm}\rule{.5mm}{0mm}%
\rule{.5mm}{.3mm}\rule{.5mm}{0mm}%
\rule{.5mm}{.3mm}\rule{.5mm}{0mm}%
\rule{.5mm}{.3mm}\rule{.5mm}{0mm}%
\rule{.5mm}{.3mm}\rule{.5mm}{0mm}%
\rule{.5mm}{.3mm}\rule{.5mm}{0mm}%
\rule{.5mm}{.3mm}\rule{.5mm}{0mm}%
\rule{.5mm}{.3mm}\rule{.5mm}{0mm} $\kappa = 0.5$\\[1ex]
\rule{2mm}{.3mm}\rule{.5mm}{0mm}\rule{.5mm}{.3mm}\rule{.5mm}{0mm}%
\rule{2mm}{.3mm}\rule{.5mm}{0mm}\rule{.5mm}{.3mm}\rule{.5mm}{0mm}%
\rule{2mm}{.3mm}
$\kappa = 0.2$\\[1ex]
\\[4.5em]
\end{minipage}\\[-4.5em]
\mycaption{Cross-sections for T-odd heavy vector boson production at the LHC as
a function of the symmetry breaking scale $f$, and for several values of the
T-odd fermions Yukawa coupling $\kappa$.}
\label{fig:vxsec}
\end{figure}

It turns out that the t-channel contributions interfere destructively with the
s-channel diagrams, as may be expected by unitarity, thus resulting in lower cross-sections for gauge boson production, see Fig.~\ref{fig:vxsec}.
For typical values $\kappa \sim {\cal O}(1)$, the LHC production cross-sections
are reduced by about one order of magnitude relative to the situation without
the t-channel diagrams ($\kappa \to \infty$). As a consequence, the expected
heavy gauge boson cross-sections are only several fb for all possible final states
$W_H^+ W_H^-$, $W_H^+ Z_H^{}$ and $W_H^+ B_H^{}$. The identification of new
physics processes of that size at the LHC relies strongly on the presence of
leptons in the signature. $W_H^\pm$ bosons decay into leptons $l=e,\mu$ with a
branching ratio of about 20\%. Folding in that branching ratio further reduces
the signal cross-section. Large backgrounds from SM gauge boson pair production
and $t\bar{t}$ production make the identification of a signal process with
${\cal O}$(fb) cross-section practically impossible.


\subsection{Mirror quarks}
\label{sc:mirrorquarks}

\begin{figure}[tb]
\vspace{1ex}
\psfig{figure=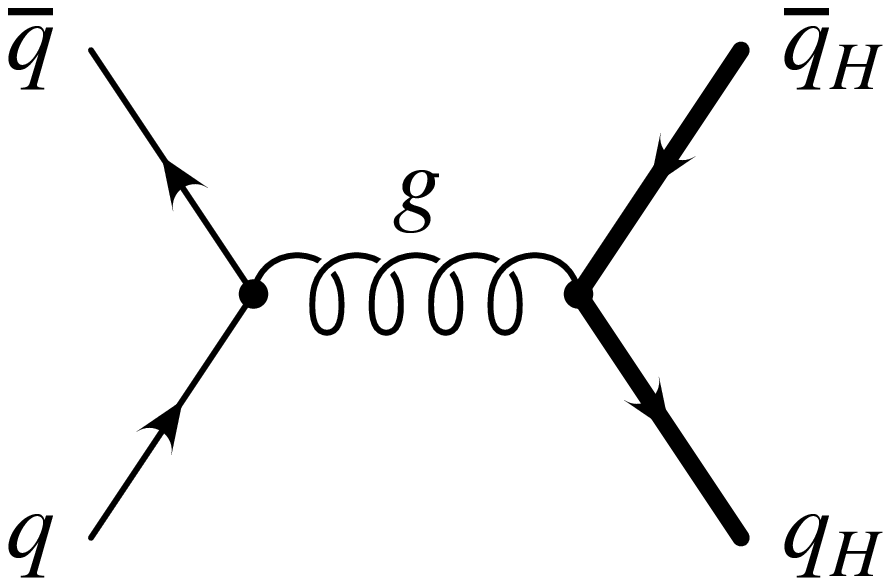, width=5cm} \hfill
\raisebox{1ex}{\begin{minipage}[b]{11cm}
\mycaption{QCD production diagram for T-odd mirror quarks at LHC. 
Thick lines indicate T-odd propagators.}
\label{fig:diaq}
\end{minipage}}
\end{figure}

\begin{figure}[tb]
\anc\hspace{1.5cm}$pp \to q_H \bar{q}_H, \; q=u,d,s,c,b$\\[-1.3em]
\anc\hspace{-4ex}%
\psfig{figure=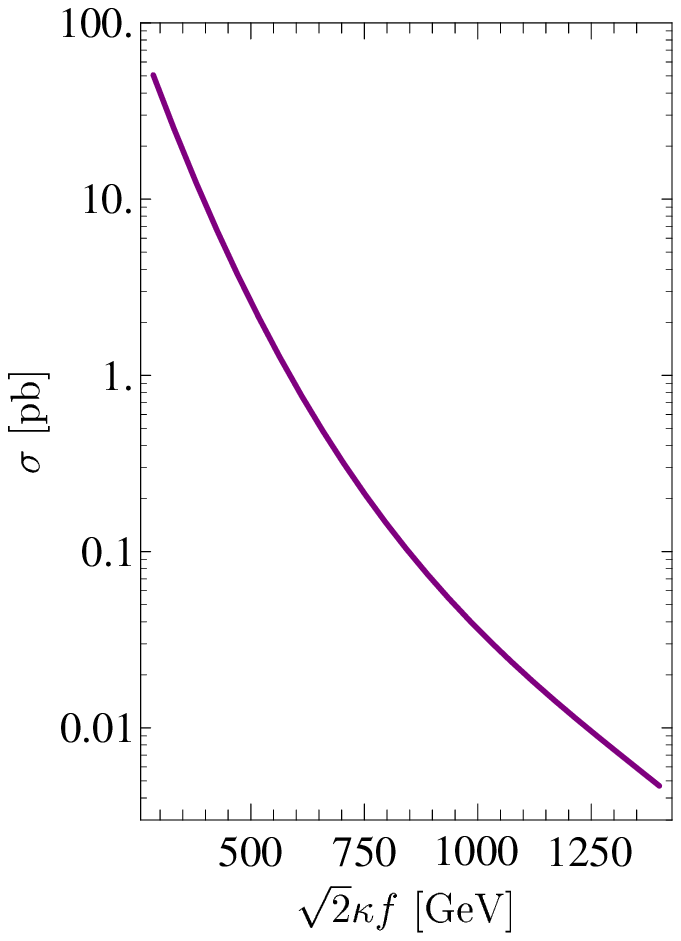, height=9.5cm} \hfill 
\raisebox{1.5cm}{\begin{minipage}{9cm}
\mycaption{Cross-sections for T-odd mirror quark production at the LHC as
a function of the quark mass $m_{q_H} \sim \sqrt{2} \kappa f$.}
\label{fig:qxsec}
\end{minipage}}
\vspace{-1em}
\end{figure}

While the mirror quarks effectively lead to a reduction of the T-odd gauge boson
production at the LHC, the mirror quarks can also be produced directly with
large cross-sections through gluon exchange, see Fig.~\ref{fig:diaq}. In
addition, there are also subdominant weak production diagrams.
Due to the strongly interacting production amplitude, the cross-section for
mirror quarks is typically larger than for T-odd vector bosons, see
Fig.~\ref{fig:qxsec}.
Since for generation- and flavor-independent $\kappa$ all mirror quarks, except
the top partners, have almost degenerate masses, the production cross-sections
are also independent of the mirror quark flavor.

For sufficiently large masses of the heavy T-odd quarks, they decay into T-odd
vector bosons. This leads to signatures that are similar to supersymmetric
theories, where squarks with large production cross-sections decay via
charginos and neutralinos. However, for a realistic assessment of the discovery
potential of the LHC for the T-odd quarks, the production rates and relevant SM
backgrounds need to be studied in more detail.

In the following the specific scenario with $f = 1$ TeV, $\kappa = 0.5$ for all
T-odd fermion flavors, and $s_\lambda = 1/\sqrt{2}$ will be considered as a
concrete example to analyze the discrimination of the mirror fermions signal
against SM backgrounds. The relevant T-odd particle masses in this scenario are
$M_{W_H^\pm} = M_{Z_H} = 647$ GeV, $M_{B_H} = 154$ GeV, $m_{q_H} = 705$ GeV for
$ q=u,d,s,c,b$ and $m_{t'_-} = 1000$ GeV.

\paragraph{(a)} {\boldmath $pp \to q_H q_H \to q q' W^\pm_H B_H \to jj\,l +
\Eslash_T$}. Here $j$ indicates a (light-flavor) jet, $l=e,\mu$ an identified
lepton, and $\Eslash_T$ stands for missing transverse energy. Since signatures
with only hadronic objects in the final state are very challenging to select
from the backgrounds, only leptonic decays of the $W_H^\pm$  are considered
here. The major SM background for the $jj\,l + \Eslash_T$ signature comes from
$pp \to W^+W^- \to q\bar{q}'l^\pm\nu_l$ production.

Here the signal and background have been calculated with CompHEP, where also
off-shell effects have been included in the computation of the
$q\bar{q}'l^\pm\nu_l$ background. Since the SM background is several orders of
magnitude larger than the signal,  the signal-to-background ratio needs to be
improved with suitable selection cuts. The result is
summarized in Tab.~\ref{tab:sel1}.
\begin{table}[tb]
\begin{tabular*}{\textwidth}{l@{\extracolsep\fill}l@{\extracolsep\fill}l@{\extracolsep\fill}l}
\hline
Cut & Signal  &
Background  & S/B \\
 & $pp \to q_H q_H \to jj\,l + \Eslash_T$ &
$pp \to q\bar{q}'l^\pm\nu_l \to jj\,l + \Eslash_T$ & \\
\hline
 & 55 fb & IR-div. & \\
\hline
$E_{T,j_1} > 200$ GeV, & & \\
$E_{T,j_2} > 50$ GeV, & & \\
$\eta_j < 2.5$,\\
 $\angle(j_1,j_2) > 30^\circ$ & 23 fb & 3.5 pb & $6.6 \times 10^{-3}$ \\
\hline
$\Eslash_T > 400$ GeV & 11 fb & 31 fb & 0.3\\
\hline
$E_{T,l} > 50$ GeV & 8 fb & 15 fb & 0.5 \\
\hline
\end{tabular*}
\mycaption{Signal and background rates for the process $pp \to q_H q_H \to q
q'\,W^\pm_H B_H \to jj\,l + /\!\!\!\!E_T$ with incremental application of signal
selection cuts. "IR-div." indicates that without any jet separation cuts, the
SM background is not infrared safe at fixed order Born approximation and thus
no number can be given here.}
\label{tab:sel1}
\end{table}
Generally, a lower threshold for the transverse energy $E_{T,j}$ of the jets
needs to be applied. Since the jet originating from the mirror squark decay into
$B_H$ is expected to be relatively hard, a rather strong cut of 200 GeV is
imposed on the transverse energy of the hardest jet. Furthermore, the jets are
required to be in the central region with rapidity $\eta_j < 2.5$ and the jets
must be separated in solid angle to be identifiable as two individual jets.
Due to the large mass of the $B_H$, which are the stable LTPs in this scenario,
the signal is characterized by large missing transverse energy$\Eslash_T$, so
that a cut on this variable is very effective to reduce the SM backgrounds.
Furthermore, since the $W$ boson from the $W_H$ decay is strongly boosted, the
final state lepton tends to be relatively hard, so that a cut on the lepton
transverse energy is useful.

We have checked that the
statistical significance of the signal cannot be improved by varying the values
of the cuts in the table. Even after the relatively aggressive
cuts in Tab.~\ref{tab:sel1}, the signal-to-background ratio is still
smaller than one, making a meaningful measurement of the T-odd production
process difficult, but not impossible. With 30 fb$^{-1}$, a statistical
significance of nine standard deviations could be achieved, but systematic
uncertainties might affect this substantially.

\paragraph{(b)} {\boldmath $pp \to q_H q_H \to q q W^+_H W^-_H \to jj\,l^+l^- +
\Eslash_T$}. With an additional lepton in the final state, this process might be
better separable from the SM background than process (a). The main SM
backgrounds are $t\bar{t}$, where both top quarks decay leptonically, and
$W^+W^-jj$, where the two jets originate from initial-state radiation.
Note that also for the signal, additional hard jets stemming from initial-state
radiation are expected \cite{susyjets}.

As before, the signal has been calculated with CompHEP, while the SM background
was simulated with MADGRAPH \cite{mad}. Again, the SM background is several
orders of magnitude larger than the signal, but can be improved with the cuts
listed in  Tab.~\ref{tab:sel2}.
\begin{table}[tb]
\begin{tabular*}{\textwidth}{l@{\extracolsep\fill}l@{\extracolsep\fill}l@{\extracolsep\fill}l@{\extracolsep\fill}l}
\hline
Cut & Signal  &
Background  & & S/B \\
 & $q_H q_H \to jj\,ll + \Eslash_T$ &
$t\bar{t} \to jj\,ll + \Eslash_T$ & $WWjj \to jj\,ll +
\Eslash_T$ \\
\hline
 & 2.7 fb \\
\hline
$E_{T,j} > 50$ GeV, \\
$E_{T,l} > 10$ GeV, \\
$\eta_j < 2.5$,\\
$\Delta R_{jj} < 0.4$ & 2.2 fb & 30 pb & 180 fb & $8 \times 10^{-5}$ \\
\hline
$\Eslash_T > 400$ GeV & 1.0 fb & 88 fb & 21 fb & 0.01\\
\hline
$E_{T,l} > 50$ GeV & 0.7 fb & 46 fb & 9 fb & 0.013 \\
\hline
b-tag veto & 0.5 fb & 3.2 fb & 7 fb & 0.03 \\
\hline
\end{tabular*}
\mycaption{Signal and background rates for the process $pp \to q_H q_H \to q q
W^+_H W^-_H \to jj\,l^+l^- + /\!\!\!\!E_T$ with incremental application 
of signal
selection cuts.}
\label{tab:sel2}
\end{table}
In addition to kinematic cuts, b-tagging also helps to reduce the $t\bar{t}$
background, since only little heavy flavor content is expected in the signal.
According to Ref.~\cite{atlas},  a b-tagging efficiency of 90\% with an
impurity of 25\% is assumed. The other selection cuts are similar to the ones
used in Tab.~\ref{tab:sel1}, with the jet cone size defined as $\Delta R_{jj} =
\sqrt{(\Delta \eta_{jj})^2 + (\Delta \phi_{jj})^2}$, where $\Delta\phi_{jj}$ is
the azimuthal angle between the two jets. It turns out that the
signal-to-background ratio remains well below one after application of the
selection cuts, and moreover the signal statistics are very low. Therefore this
channel does not look promising as a discovery mode for the LHT model at the
LHC.

\paragraph{(c)} {\boldmath $pp \to q_H q_H \to q q' W^\pm_H Z_H \to  q q' W^\pm
h^0 B_H B_H \to jj\,bb\,l + \Eslash_T$}. Since the $Z_H$ boson almost always
decays into the (little) Higgs boson $h^0$, the selection of this signal
process needs to make use of the two b jets in the final state. The largest
SM background is semileptonic $t\bar{t}$ decays.

Again, CompHEP was used to calculate the signal, while the SM background
was simulated with MADGRAPH. For the signal selection procedure, it is assumed
that the Higgs boson has already been discovered and its mass measured, so that
a cut can be applied on the invariant mass $m_{\rm bb}$ of the two bottom jets
stemming from the Higgs. This requires a good identification of the b jets among
the four or more jets in the event. Thus b-tagging is mandatory for the signal
selection, and it is advantageous to optimize the b-tagging procedure for a high
purity of 98\%, at an efficiency of 60\% \cite{atlas}.
Applying the cuts in Tab.~\ref{tab:sel3},
\begin{table}[tb]
\begin{tabular*}{\textwidth}{l@{\extracolsep\fill}l@{\extracolsep\fill}l@{\extracolsep\fill}l}
\hline
Cut & Signal  &
Background  & S/B \\
 & $q_H q_H \to jj\,bb\,l + \Eslash_T$ &
$t\bar{t} \to jj\,bb\,l + \Eslash_T$ \\
\hline
 & 14.0 fb \\
\hline
$E_{T,j} > 50$ GeV, \\
$E_{T,l} > 10$ GeV, \\
$\eta_j < 2.5$,
$\Delta R_{jj} < 0.4$ & 13.4 fb & 128 pb & $1 \times 10^{-4}$ \\
\hline
$\Eslash_T > 250$ GeV & 9.8 fb & 3.1 pb & $3 \times 10^{-3}$\\
\hline
$E_{T,l} > 50$ GeV & 8.4 fb & 1.3 pb & $6 \times 10^{-3}$ \\
\hline
b-tag & 5.0 fb & 770 fb & $6 \times 10^{-3}$ \\
\hline
$100 \gev < m_{\rm bb} < 150 \gev$ & 5.0 fb & 65 fb & 0.08 \\
\hline
\end{tabular*}
\mycaption{Signal and background rates for the process $pp \to q_H q_H \to q q'
W^\pm_H Z_H \to jj\,bb\,l + /\!\!\!\!E_T$ with incremental application 
of signal
selection cuts.}
\label{tab:sel3}
\end{table}
it is found that the signal-to-background ratio stays well below one after
application of the selection cuts, so that this channel is also not suitable
for new physics discovery in the LHT model.

\paragraph{(d)} {\boldmath $pp \to t'_- \bar{t}'_- \to t \bar{t} B_H B_H$}. 
The final state signature of this process is identical to $t\bar{t}$ production.
As an example, the semileptonic decay of the top quark pair is considered,
leading to the final state $jj \, bb \, l + \Eslash_T$.

Signal and background have been computed as above, using b-tagging with
98\% purity and 60\% efficiency. The main discrimination to the $t\bar{t}$
background is a cut on missing transverse energy, see Tab.~\ref{tab:sel4},
\begin{table}[tb]
\begin{tabular*}{\textwidth}{l@{\extracolsep\fill}l@{\extracolsep\fill}l@{\extracolsep\fill}l}
\hline
Cut & Signal  &
Background  & S/B \\
 & $t'_- \bar{t}'_- \to jj\,bb\,l + \Eslash_T$ &
$t\bar{t} \to jj\,bb\,l + \Eslash_T$ \\
\hline
$E_{T,j} > 50$ GeV, \\
$E_{T,l} > 10$ GeV, \\
$\eta_j < 2.5$,
$\Delta R_{jj} < 0.4$ & 8.2 fb & 128 pb & $6 \times 10^{-5}$ \\
\hline
$\Eslash_T > 800$ GeV & 1.4 fb & 3.2 fb & $0.4$\\
\hline
b-tag & 0.8 fb & 1.9 fb & 0.4 \\
\hline
\end{tabular*}
\mycaption{Signal and background rates for the process $pp \to t'_- \bar{t}'_-
\to t \bar{t} B_H B_H \to jj \, bb \, l + /\!\!\!\!E_T$ with incremental application 
of signal selection cuts.}
\label{tab:sel4}
\end{table}
which improves the signal-to-background ratio tremendously. However, the
resulting signal-to-background ratio is still below one, and the remaining
signal rate is small, so that even without considering systematic uncertainties,
only a statistical significance of less than three standard deviations is 
achievable with 30 fb$^{-1}$.

\paragraph{}Other decay chains, such as $pp \to q_H q_H \to q q' Z_H Z_H$ have large SM
background and will not be considered.


\section{Mirror leptons and decay signatures}
\label{sc:sign}

Mirror leptons, the T-odd partners of the leptons, can be produced directly at
the LHC through s-channel exchange of SM gauge bosons. However, their production
cross-sections are small, below 1 fb
for the scenario with $\kappa_l = 0.5$ and $f = 1000$ GeV, i.e. $m_{l_H} = 707$
GeV. We therefore do not consider the direct production further.

However, the mirror leptons may play an important role in the decay signatures.
If the mirror fermions are not degenerate, but for instance T-odd leptons and
T-odd quarks have different masses, the experimental signatures can be greatly
altered. For example for $\kappa_q = 0.5$ and $\kappa_l = 0.2$, the mirror
lepton mass is $m_{l_H} = 283$ GeV, and the heavy gauge
boson can decay into the mirror leptons, $Z_H \to l^\pm l^\mp_H$. If this decay
channel is open, the branching ratio will be almost 100\% for all lepton
flavors combined. As a consequence, the T-odd quarks
can decay through cascades like 
\begin{equation}
q_H \to q \, Z_H \to q \, l^\pm l^\mp_H \to
q \, l^+l^- \, B_H, \label{eq:samel}
\end{equation}
leading to a signal of opposite-sign same-flavor leptons
and missing transverse energy, similar to the situation in mSUGRA scenarios in
supersymmetry \cite{msugralhc}. Here one can make use of the fact that the main
SM backgrounds produces uncorrelated leptons, with the same proportion of
same-flavor ($e^+e^-$/$\mu^+\mu^-$) and opposite-flavor ($e^\pm\mu^\mp$)
leptons. If then the opposite-flavor events are subtracted from the total
sample, the SM backgrounds are effectively removed, while the little Higgs
signal is not affected \cite{msugralhc}. However, the statistical noise from the
backgrounds is not reduced by this method and still affects the extraction of
the signal process.

For the signal selection of the decay chain eq.~\eqref{eq:samel}, the same cuts
as listed in Tab.~\ref{tab:sel2} can be used. Since the decay of one mirror
quark already leads to two leptons in the final state, no assumption for the
decay of the second mirror quark needs to be made, other than that it leads to
the LTP $B_H$ in the final state, which generates missing transverse energy in
the signature. This improves the signal statistics compared to the analysis in
section~\ref{sc:mirrorquarks}~(b). After application of the cuts in
Tab.~\ref{tab:sel2}, the signal rate for the process eq.~\eqref{eq:samel} is 
21 fb, while the SM backgrounds amount to 10 fb.
If in addition the different-lepton-flavor subtraction described above is used,
the remaining backgrounds are negligible, but introduce some statistical noise.
Combining the statistical errors, the new physics signal can be identified in
this channel with more than 20 standard deviations with 30 fb$^{-1}$ luminosity.

\begin{figure}[tb]
\anc\hspace{1.5cm}
\psfig{figure=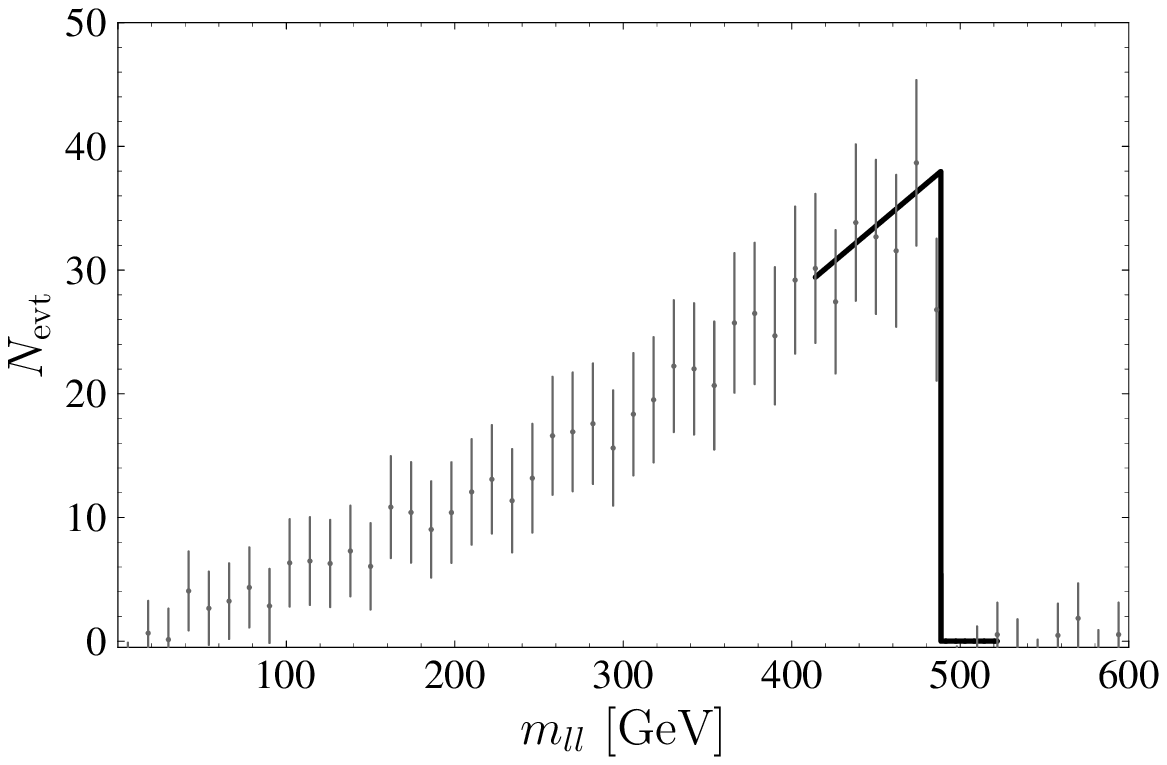, width=12cm}
\mycaption{Distribution of the di-lepton invariant mass $m_{ll}$ associated with
the decay chain $q_H \to q \, Z_H \to q \, l^\pm l^\mp_H \to
q \, l^+l^- \, B_H$, for the parameter values $f =1000 \gev$, $\kappa_q = 0.5$ 
and $\kappa_l = 0.2$. The error bars correspond to the statistical uncertainty
for 30 fb$^{-1}$ luminosity, after subtraction of backgrounds.}
\label{fig:mll}
\end{figure}

Furthermore, the di-lepton invariant mass distribution shown in
Fig.~\ref{fig:mll} exhibits a distinct upper endpoint
at
\begin{equation}
(m_{ll}^{\rm max})^2 = (m^2_{Z_H} - m^2_{l_H})(m^2_{l_H}-m^2_{B_H})/m^2_{l_H}.
\label{eq:mllmax}
\end{equation}
This feature can be used to extract some
information about the T-odd particle masses. The spectrum
edge can be fitted with a simple triangle-shaped fit function, see
Fig.~\ref{fig:mll}.
Assuming 30 fb$^{-1}$ luminosity, one obtains
\begin{equation}
m_{ll}^{\rm max} = 488.6^{+5.2}_{-4.4} \gev,
\end{equation}
which is in good agreement with the input value of the underlying model, 
488.5~GeV. Due to the relatively small signal cross-section, 
the statistical error is quite large. It can be improved by using more
luminosity. With a total luminosity of 300 fb$^{-1}$, the error is reduced
to 
\begin{equation}
\delta m_{ll}^{\rm max} = \ ^{+2.1}_{-1.7} \gev.
\end{equation}
Still, at this level of precision, systematic errors due to the lepton energy
scale uncertainty or due to mistagging can be neglected. 

Since eq.~\eqref{eq:mllmax} depends only mildly on $m_{l_H}$ for $m_{B_H} \ll
m_{l_H} \ll m_{Z_H}$, the measured endpoint $m_{ll}^{\rm max}$ gives a rough
estimate of the mass difference between the T-odd gauge bosons, $m_{ll}^{\rm
max} \approx m_{Z_H}-m_{B_H}$. In the case of the scenario studied here,
$m_{Z_H}-m_{B_H} = 493$ GeV,  so that this simplified relation holds within
statistical errors.

More information about
the mass spectrum of the T-odd particle could be obtained by studying invariant
mass distributions including the jet originating from the mirror quark decay
\cite{msugralhc}. However, due to additional jet radiation and ambiguities in
the selection of the jet, the kinematic endpoints of these distributions are
more smeared out. This fact together with the low signal statistics leads to
very large errors for fits to kinematic endpoints of jet-lepton distributions.


\section{Conclusion}
\label{sc:concl}

In this paper we have considered the production and decay of the T-odd heavy
particles at LHC in littlest Higgs models with T-parity. This symmetry implies
the presence of heavy mirror particles and that the heavy particles  can only
be pair produced. In the case of heavy gauge bosons, 
s-channel and t-channel production mechanisms interfere destructively. This
reduces the production rate substantially to less than about $0.1$ pb and makes
this channel  almost impossible to observe. 

Similarly, the discovery and measurement of T-odd mirror quarks at the LHC is
very difficult. We have considered all possible decay chains, but  most decay
channels are lost in the SM background. Only the channel $pp \to q_H q_H \to q
q' W^\pm_H B_H \to jj\,l + \Eslash_T$ with  a production rate of about 10 fb
might be promising, but mandates further study. We discuss to some extent the
cuts required to reduce the SM background (see table \ref{tab:sel1}).  Although
the final state signatures are quite similar to the ones for squark production
in the MSSM, the signal rates are typically lower in the LHT models than in the
MSSM because there is no partner to the gluon. Recall that  in supersymmetric
models, the partner of the gluon, the gluino, is typically the primary particle
for squark production. This yields larger cross-section and additional hard
jets in the final state, which help to discriminate from the background. In LHT
models however, the gluon is a has no partner. It should be noted, however,
that the somewhat pessimistic results of this section strongly depend on the
underlying scenario. For example, by lowering the breaking scale $f$, the
production cross-section of T-odd quarks at the LHC are greatly enhanced.
Fig.\ref{fig:qxsec} gives a rough picture of the dependence on $f$; the actual
dependence is more complicated because of the intricate energy dependence of
the cuts.

Mirror leptons, are only produced through weak processes, but may play a role
in the decay  signatures of the heavy quarks if they are  lighter than the
heavy quarks or gauge bosons. Then the leptonic branching ratio of the neutral
gauge boson $Z_H$ is close to one, with opposite-sign same-flavor leptons. As
detailed above, this can be used to effectively suppress the SM background;
although some non-negligible statistical noise remains. Moreover, this decay
could also yield information about the spectrum of the T-odd particles. Clearly
detailed experimental studies along these lines will be required for a more
conclusive assessment.

\bigskip

\vspace{- .3 cm}
\section*{Acknowledgments}
This work was supported by the Schweizer Nationalfonds.


\end{document}